\documentclass[12pt]{article}
\usepackage{epsf}
\usepackage{amsmath}
\def\be{\begin{equation}}
\def\ee{\end{equation}}

\begin{document}
\begin{titlepage}
\begin{center}
{\Large \bf William I. Fine Theoretical Physics Institute \\
University of Minnesota \\}
\end{center}
\vspace{0.2in}
\begin{flushright}
FTPI-MINN-08/36 \\
UMN-TH-2718/08 \\
September 2008 \\
\end{flushright}
\vspace{0.3in}
\begin{center}
{\Large \bf Diresonance in production and scattering
of heavy mesons
\\}
\vspace{0.2in}
{\bf S. Dubynskiy \\}
School of Physics and Astronomy, University of Minnesota, \\ Minneapolis, MN
55455, USA, \\

{\bf M.B. Voloshin  \\ } William I. Fine Theoretical Physics
Institute, University of
Minnesota,\\ Minneapolis, MN 55455, USA \\
and \\
Institute of Theoretical and Experimental Physics, Moscow, 117218, Russia
\\[0.2in]
\end{center}

\begin{abstract}

We consider the production and scattering amplitudes of heavy mesons in a
situation, where there are two closely spaced narrow resonances, which structure
we refer to as a diresonance. Assuming strong overlapping of the resonances
coupled to common channels, it is found, using the unitarity and analyticity
constraints, that the production amplitudes by a weak source should have similar
behavior with energy in different channels. In particular the ratio of the
coefficients for each pole contribution to the production amplitude is fixed at
$-1$.

\end{abstract}

\end{titlepage}

The spectroscopy of resonances near the charm threshold attracts a considerable
renewed interest. Recent experimental data not only provide evidence of new
states but also suggest that the nature of well known resonances merits a
reexamination at the level of fine details. The new exotic charmonium-like
states\cite{go} considerably expand the standard spectrum of charmonium, and
challenge us for a better understanding of the strong dynamics. Furthermore the
apparent significance of multiquark states at the onset of open charm threshold
may
also impact the properties of the known resonances and may be instrumental in
resolving some long-standing puzzles. One such puzzle is related to an
inconsistency between BES and CLEO results for the production and decays of
$\psi (3770)$ resonance, in particular the fraction of its decays into non-$D
\bar D$ states\cite{egmr,mv}. Recently BES Collaboration has reanalyzed their
data on $e^+e^-$ annihilation in the energy region between 3.700 and 3.872 GeV
\cite{Ablikim:2008et}. They reported observation of an anomalous line-shape
behavior of the cross section which is inconsistent with the presence of only
one simple $\psi (3770)$ resonance in this energy region. It is claimed that
this anomalous behavior could be better understood in terms of two resonances
near the c.m. energies of 3.764 GeV and 3.779 GeV. This result could violate the
conventional interpretation of  $\psi (3770)$ as being a dominantly $1\, ^3D_1$
charmonium state with an admixture of $2\, ^3 S_1$, and clearly suggests a more
complicated structure, possibly including strong dynamics of the $D$ meson pairs
near the threshold. Namely, a diresonance structure may arise from existence of
both a charmonium state and a `molecular' $D \bar D$ threshold resonance.

Apriori one would expect that the nature of each of the two individual peaks
could be studied by further exploring their relative coupling to various
channels. However the purpose of the present paper is to argue that this
standard method, applicable to sufficiently widely separated resonances, is
unlikely to be applicable for a strongly overlapping pair of resonances, such as
the one indicated by the BES data, i.e. when the splitting between the positions
of the resonances is comparable with their widths, and all of these parameters
are small in a typical energy scale for the process. (It is natural to call such
a structure as diresonance.) Using unitarity and analyticity constraints, we
find, under the simplest assumptions, similar to those involved in the standard
Breit-Wigner treatment of a single resonance, that the two states are
necessarily strongly mixed, and that various final states produced e.g. in the
$e^+e^-$ annihilation have the behavior of the production cross section in the
diresonance region proportional to one another. Furthermore, if a diresonant
production amplitude is written as a linear combination of two poles, then in
the limit where the diresonance parameters, the two widths and the splitting
between the poles, can be considered as small, the relative factor between the
two poles is necessarily equal to $-1$.

In practical terms this universal behavior implies that it would be problematic
to disentangle experimentally the underlying origin of the states in a
diresonance complex. In particular, even if the suggested by the experiment
structure around 3.77 GeV, originates from an overlap of charmonium and
molecular states, the mixing between them effectively erases any difference in
their experimental signature.

We start  our discussion of a diresonance structure with a simple case of just
one scattering channel.  The radial part of the wave function of the
particle with mass $m$ with arbitrary
complex energy $E$ and orbital momentum $l=0$ (here we consider
$S$-wave motion for simplicity) has the following form at large distances $r$:
\be
R={1\over
r}\,\left[B^*(E)\,e^{i\,kr}+B(E)\,e^{-i\,kr}\right]\,,\,\,\,k=\sqrt{2
m E}\,,
\label{swf}
\ee
where the coefficients of the incoming and outgoing waves are related by complex
conjugation due to the requirement that the wave function is real at real
negative $E$.
In the familiar case of a single resonance at
$E_r=E_0-i\,\Gamma/2$, with $E_0$ and $\Gamma$ being the position and
the width of the resonance, one has $B(E_r)=0$, which ensures that the wave
function of the resonant state
vanishes at spatial infinity\cite{ll}. One can then expand the
function $B(E)$ near the position of the resonant level $E_r$ as
\be
B(E)=(E-E_0+{i \over 2}\,\Gamma)\,b\,,
\label{bofe1}
\ee
with $b$ being a smooth function of energy, i.e. $b$ changes on a scale much
larger than the resonance width $\Gamma$. The scattering
$S$-matrix element is then found as follows
\be
S=\exp (2 i\,\delta)={B^*(E)\over B(E)}={E-E_0-i\,\Gamma/2\over
E-E_0+i\,\Gamma/2}\,\,\exp (2 i\,\delta^{(0)})\,,
\ee
where  $\delta^{(0)}$ is a nonresonant phase which is a smooth
function of energy defined as $\exp(2i\delta^{(0)})=b^*/b$.

This standard Breit-Wigner treatment of a single resonance can be readily
extended to the case of a diresonance, i.e. in the situation when the scattering
amplitude has two closely separated poles at $E_1-i \Gamma_1/2$ and $E_2-i
\Gamma_2/2$ with both widths $\Gamma_1$ and $\Gamma_2$ and the difference
$E_1-E_2$ being considered as `small'. The expansion of the coefficient $B(E)$
having two zeros in the diresonance region is obviously given by
\be
B(E)=(\Delta_1 + i \gamma_1) (\Delta_2 + i \gamma_2) \, b~,
\label{bofe2}
\ee
where the notation is introduced $\Delta_a= E-E_a$, $\gamma_a = \Gamma_a/2$
($a=1,\,2$), and, similarly to Eq.(\ref{bofe1}), the coefficient $b$ is a slowly
varying function of the energy, which can be approximated by a constant on the
energy scale of the diresonance region. The corresponding expression for the $S$
matrix element then takes the form
\be
S={(\Delta_1-i\,\gamma_1)
(\Delta_2-i\,\gamma_2) \over
(\Delta_1+i\,\gamma_1)(\Delta_2+i\,\gamma_2)} \, \exp (2 i\,\delta^{(0)}) \,,
\ee

Let us consider now the amplitude $A(E)$ for production of the scattering state
in the diresonance region by a point-like source. The production process is
assumed to be weak, so that it is sufficient to consider only the lowest order
in the coupling to the source. The energy dependence of such amplitude is
proportional to the inverse of the coefficient of the incoming wave in the wave
function (\ref{swf}): $A(E) = g/B(E)$ with $g$ being a real (at real $E$) smooth
function of energy. Indeed, according to the familiar ``$\psi(0)$ rule" the
absolute value of the amplitude is proportional to $\psi(0)$, provided that the
wave function is normalized to a fixed amplitude at infinity, $R = (1/r) \,
\sin(kr + \delta)$, which implies the relation $|A(E)| \propto 1/|B(E)|$. On the
other hand, according to the Watson's theorem, the phase of $A$ is given by
$\delta$, i.e. the phase is that of $1/B(E)$. Using this relation we readily
find an analytical formula for the production amplitude in the diresonance
energy region
\be
A(E)={(g / b) \over  (\Delta_1+i\,\gamma_1)(\Delta_2+i\,\gamma_2)}~.
\label{aofe}
\ee

The latter expression for the diresonance production amplitude when written as a
sum over two resonances:
\be
A(E)={(g / b) \over E_1-E_2+ i \gamma_2-i \gamma_1 } \, \left ( {1 \over
\Delta_1+i\,\gamma_1} -  {1 \over \Delta_2+i\,\gamma_2} \, \right )
\label{a1res}
\ee
tells us that the relative phase between the two resonance factors has to be
equal to $\pi$ and the coefficients of the pole factors should be the same.

As a simple cross check we considered a toy model with the scattering of a
particle with mass $m$ in a central potential with two Gaussian barriers:
\be
V(r)={1 \over 2 \, m \, r_0^2} \, x \, \left \{ h_1\,
\exp \left [ -{(x-x_1)^2 \over w_1} \right ]+h_2\, \exp \left [ -{(x-x_2)^2
\over w_2} \right ] \right \}~,
\label{mpot}
\ee
where $x=r/r_0$ is a dimensionless ratio of the distance $r$ to an arbitrary
scale $r_0$ and the parameters $h_i,x_i,w_i$ are also dimensionless. The 
distance scale $r_0$ also sets the scale $E_0=(2 m r_0^2)^{-1}$ for the energy
of the particle.
We calculated numerically the  dimensionless ``production amplitude" as the
value of the wave
function at the origin, $\psi(0)$, at energy $E$, provided that at large $r$ the
wave function is normalized to a wave with unit amplitude. We found that every
time the parameters of the potential $h_i,x_i,w_i$ are tuned in such a way that
a diresonance structure appears, the production amplitude is closely
approximated by the expression (\ref{a1res}). We illustrate this behavior in
Fig.~1 for one specific set of parameters ($h_1=5.5, h_2=0.94, x_1=1.75,
x_2=6.92, w_1=0.86, w_2=0.45$). The fit curve shown in the plot corresponds to a
constant ratio $g/b$ in Eq.(\ref{a1res}), so that the relative factor between
the two poles  exactly equals $-1$. If the fit is relaxed and this relative
factor is also treated as a fit parameter, we find the best approximation for it
as $-0.90+0.02 i$.

\begin{figure}[ht]
\begin{center}
 \leavevmode
    \epsfxsize=10cm
    \epsfbox{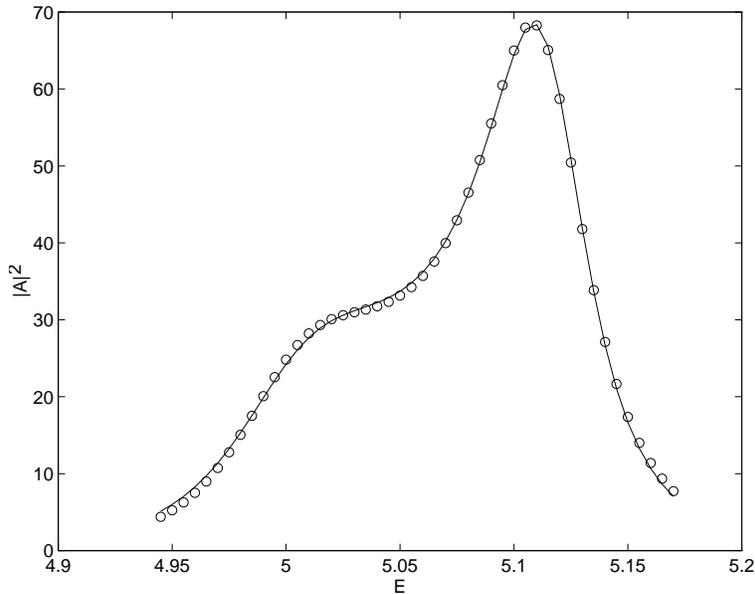}
    \caption{The  dimensionless ``production cross section" $|A(E)|^2$ vs the
energy $E$ (in units of $E_0$) in
the diresonance region in a toy model with potential scattering. The circles are
the numerical data and the curve is the fit by the formula in Eq.(\ref{a1res}).
}
\end{center}
\end{figure}

It may appear that the rigid constraint on the relative contribution of the two
single-resonance factors in a diresonance complex is a limitation of the
considered situation with one scattering channel. For this reason we proceed to
discussing scattering in two coupled channels, where we find that the same
constraint on the relative strength of the two pole factors applies in each of
the channels, so that in fact the production of the final states in the two
channels is described by the same energy dependence with the only free factor
being the ratio of the overall yield of each of the final channels.

In this discussion we consider two channels $a$ and $b$ and consider the $S$
matrix for the scattering and the production by a localized source:
\be
S=\begin{pmatrix}
  S_{aa} & S_{ab} &i A_a \\
   S_{ba}& S_{bb}  & i A_b\\
   i A_a  & i A_b  &1 \\
\end{pmatrix}~,
\label{sdef}
\ee
where $A_{a,b}$ are the production amplitudes for each of the channels by a weak
localized source, so that the quadratic in the strength effect of the source in
the diagonal (33) element of the $S$ matrix can be neglected.  In the zeroth
order in the production amplitudes we find that the general solution to the
unitarity conditions for the two-channel scattering matrix with a diresonance
singularity has the form
\be
S_{m n}=\left [\, \delta_{m n}-{2 i\, (\Delta_1\,\gamma_2+\Delta_2\gamma_1)\,
\eta_m \eta_n\over
(\Delta_1+i\,\gamma_1)(\Delta_2+i\,\gamma_2)} \right ] \exp [ i (\delta_m^{(0)}+ \delta_n^{(0)} )]\,;\,\,\,\,m,\,
n=a,\,b\,,
\label{smn}
\ee
where $\delta_{a,b}^{(0)}$ is the nonresonant scattering phase in the corresponding channel and the real factors $\eta_a$ and $\eta_b$ satisfy the
condition $\eta_a^2+\eta_b^2=1$. It can be noted that in the single resonance
case these factors are determined by the corresponding branching fractions for
the resonance: $\eta_n^2=\Gamma_n/\Gamma$. In the diresonance case we do
not find a simple direct relation of these factors to the individual width
parameters $\Gamma_1$ and $\Gamma_2$. However these factors can still be interpreted in terms of the branching ratios (for the diresonance complex) in the sense that $\eta_n^2$ gives the probability of the branching of the scattering into the corresponding channel. It is also quite clear that setting $\eta$
to zero in one channel and $\eta^2=1$ in the other, returns us to the previously
discussed case of a single channel.

The unitarity relation for the matrix (\ref{sdef}) in the first order in the
production amplitudes $A_{a,b}$ then yields these amplitudes in the form
\be
A_n={\mu \, \eta_n \over
(\Delta_1+i\,\gamma_1)(\Delta_2+i\,\gamma_2)}\,\exp(i \delta_n^{(0)} );\,\,\,n=a,\,b\,,
\label{anres}
\ee
where the real smooth factor $\mu$ characterizes the strength of the coupling
of the source.

The expression (\ref{anres}) implies that the single-channel behavior of a
diresonance production amplitude also holds for multiple channels. Namely, when
expressed as a linear combination of two poles the relative coefficient between
the two pole terms is necessarily equal to $-1$ as shown in Eq.(\ref{a1res}), and the relative yield in each channel is determined by the branching factor $\eta^2$.

One can readily notice that the relative factor $-1$ is a trivial consequence of
the production amplitude behaving as $(\Delta E)^{-2}$ away from the diresonance
region. Clearly, in order to invalidate such a behavior, one would have to
introduce in the dependence of the coefficient $g/b$ an energy scale comparable
to the parameters $(E_1-E_2)$, $\Gamma_1$ and $\Gamma_2$. Thus our conclusions
are applicable in the situation, which we refer to as a diresonance, where these
parameters are small in the typical scale in the problem. In this respect the
assumption is quite similar to the familiar Breit-Wigner approximation for a
single resonance, where the resonance has to be considered as narrow, i.e. with
a small width. In the diresonance case it is also the splitting between the
poles, which has to be ``narrow" in addition to the width parameters, for our
approximation to be valid.

We believe that our consideration of a diresonance structure is quite generic
and may be applicable to the suggested by experiment\cite{Ablikim:2008et}
structure in the $e^+e^-$ cross section near 3.77 GeV, and possibly to other
similar structures. A specific detailed application of the discussed diresonance
properties to the production of $D$ meson pairs in the $\psi(3770)$ region
should also include the $P$-wave kinematics with different thresholds for the
pairs of neutral and charged mesons, as well as the Coulomb effects. Such
analysis can be done along the lines presented in Ref.~\cite{5auth}, if more
detailed data become available. At this point we can only remark that the
two-pole fit to the data\cite{Ablikim:2008et} does not contradict the expression
(\ref{a1res}) for the production amplitude. However the error range is still too
large for any further conclusions to be drawn.

This work is supported in part by the DOE grant DE-FG02-94ER40823. The work of
S.D. is supported in part by the Stanwood Johnston grant from the Graduate
School of the University of Minnesota.

\end{document}